\documentclass[natbib]{svjour3_mod}             

\smartqed  

\usepackage{graphicx}
\usepackage{amsmath}
\usepackage{amssymb}

\newcommand{\earth}{\oplus}

\newcommand{\aap}{{A\&A}}	
\newcommand{\nat}{{Nature}}	
\newcommand{\icarus}{{Icarus}}
  
\newcommand{\aj}{{AJ}}	
\newcommand{\apj}{{ApJ}}	
\newcommand{\apjl}{{ApJ Lett.}}
\newcommand{\mnras}{{MNRAS}}	
	
\newcommand{\planss}{{Planet.~Space~Sci.}}	
	
\newcommand{\gca}{{Geochim.~Cosmochim.~Acta}}	
\newcommand{\ssr}{{Space~Sci.~Rev.}}
\newcommand{\araa}{{ARA\&A}}

\newcommand{\doiurl}{https://doi.org/}

\begin{document}

\title{The Delivery of Water During Terrestrial Planet Formation}


\author{David P.~O'Brien \and Andre Izidoro \and Seth A.~Jacobson \and Sean N.~Raymond \and David C.~Rubie}

\institute{D.~P.~O'Brien \at
Planetary Science Institute \\
1700 E.~Ft.~Lowell, Suite 106 \\
Tucson, AZ 85719, USA \\
\email{obrien@psi.edu}
\and
A.~Izidoro (current institution) \at
UNESP, Universidade Estadual Paulista \\
Grupo de Dinamica Orbital Planetologia \\
Guaratingueta, CEP 12.516-410, Sao Paulo, Brazil
\and
A.~Izidoro (institution at time of submission) \at
Laboratoire d'Astrophysique de Bordeaux \\
Universite de Bordeaux, CNRS, B18N \\
Allee Geoffroy Saint-Hilaire \\
33615 Pessac, France
\and
S.~A.~Jacobson (current institution) \at
Department of Earth and Planetary Sciences \\
Northwestern University \\
Evanston, IL 60208, USA
\and
S.~A.~Jacobson (institution 1 at time of submission) \at
Bayerisches Geoinstitut \\
University of Bayreuth \\
D-95490 Bayreuth, Germany
\and
S.~A.~Jacobson (institution 2 at time of submission) \at
Observatoire de la C\^ote d'Azur \\
Bd. de l'Observatoire \\
CS 34229, F-06304, Nice Cedex 4, France
\and
S.~N.~Raymond \at
Laboratoire d'Astrophysique de Bordeaux \\
Universite de Bordeaux, CNRS, B18N \\
Allee Geoffroy Saint-Hilaire \\
33615 Pessac, France 
\and
D.~C.~Rubie \at 
Bayerisches Geoinstitut \\
University of Bayreuth \\
D-95490 Bayreuth, Germany}

\journalname{Space Science Reviews}

\date{Accepted Jan 17, 2018 for publication in \textit{Space Science Reviews} under the topical collection \textit{The Delivery of Water to Protoplanets, Planets and Satellites} and in an ISSI Space Science Series book of the same title.}

\maketitle

\begin{abstract}

The planetary building blocks that formed in the terrestrial planet region were likely very dry, yet water is comparatively abundant on Earth.  Here we review the various mechanisms proposed for the origin of water on the terrestrial planets.  Various in-situ mechanisms have been suggested, which allow for the incorporation of water into the local planetesimals in the terrestrial planet region or into the planets themselves from local sources, although all of those mechanisms have difficulties.  Comets have also been proposed as a source, although there may be problems fitting isotopic constraints, and the delivery efficiency is very low, such that it may be difficult to deliver even a single Earth ocean of water this way.  The most promising route for water delivery is the accretion of material from beyond the snow line, similar to carbonaceous chondrites, that is scattered into the terrestrial planet region as the planets are growing.  Two main scenarios are discussed in detail.  First is the classical scenario in which the giant planets begin roughly in their final locations and the disk of planetesimals and embryos in the terrestrial planet region extends all the way into the outer asteroid belt region.  Second is the Grand Tack scenario, where early inward and outward migration of the giant planets implants material from beyond the snow line into the asteroid belt and terrestrial planet region, where it can be accreted by the growing planets.  Sufficient water is delivered to the terrestrial planets in both scenarios.  While the Grand Tack scenario provides a better fit to most constraints, namely the small mass of Mars, planets may form too fast in the nominal case discussed here.  This discrepancy may be reduced as a wider range of initial conditions is explored.  Finally, we discuss several more recent models that may have important implications for water delivery to the terrestrial planets.


\end{abstract}


\section{Introduction}
\label{sec:intro}

The existence of water on the Earth is somewhat of a puzzle.  Evidence suggests that solids that condensed around 1 AU in the solar nebula were dry and highly reduced, yet Earth clearly has water on its surface, and likely even more in its mantle.  Other terrestrial planets show evidence for water as well.  The large D/H ratio of Venus' atmosphere suggests an ancient ocean was lost to space \citep{Donahue1982Sci}, while geomorphology \citep{Baker2007Elem}, neutron spectroscopy data \citep{Feldman2004JGR}, and various other lines of evidence point to a wet past for Mars and significant water still present as subsurface ice.  Numerous mechanisms have been suggested for delivering this water to the terrestrial planets, or allowing for incorporation from local sources.  In this chapter, we review the different mechanisms and assess their plausibility, in light of available constraints and evidence. 

We begin the chapter with an overview of the key geochemical and cosmochemical evidence for the origin of Earth's water in Section \ref{sec:constraints}.  In Section \ref{sec:stages} we describe the different stages of terrestrial planet formation, from the growth of planetesimals and planetary embryos through the final stage of accretion, and discuss two different scenarios for late-stage accretion.  We first describe the `classical' scenario, with a disk of planetesimals and embryos extending from the terrestrial planet region into the outer asteroid belt region, and the giant planets in roughly their current locations \citep[e.g.][]{Chambers2001Icar, Raymond2004Icar, OBrien2006bIcar, Raymond2009Icar, Izidoro2013ApJ, Izidoro2014ApJ}.  Second, we discuss the `Grand Tack' scenario of \citet{Walsh2011Nat}, where the giant planets experience an inwards-then-outward migration that implants material from beyond the snow line into the asteroid belt and terrestrial planet region.

Beginning our discussion of water delivery, we first describe various `in-situ' mechanisms in Section \ref{sec:insitu} that could allow for the incorporation of water into planetesimals local to the growing terrestrial planets or directly into the planets themselves.  We then discuss the possibility of water delivery by comets in Section \ref{sec:comets}.  Problems and difficulties with both scenarios are discussed.  Finally, in Section \ref{sec:latedeliv} we discuss water delivery in the classical and Grand Tack scenarios of late-stage planet formation.  Both scenarios allow for sufficient water delivery to explain the water budget of Earth.  The Grand Tack scenario provides the best fit to most constraints, namely the small mass of Mars, but some issues with the timescale of planet formation remain.  We summarize our discussion and point to further avenues of research on this topic in Section \ref{sec:summary}.

\section{Geochemical and Cosmochemical Constraints}
\label{sec:constraints}

Here we discuss some of the key constraints on modeling of water delivery to the terrestrial planets, with a focus on the Earth.  For further details see the other chapters in this collection by \citet{Peslier2017SSRv} on water in the Earth and \citet{Alexander2018SSRv} on water in small Solar System bodies.

\citet{Lecuyer1998ChemGeo} estimate the total mass of water in the Earth's crust, oceans, and atmosphere to be $2.8\times10^{-4}$ $\mathrm{M_{\earth}}$.  The water content of the mantle is more uncertain, with estimates ranging from $0.8-8\times10^{-4}$ $\mathrm{M_{\earth}}$ \citep{Lecuyer1998ChemGeo} to $2 \times10^{-3}$ $\mathrm{M_{\earth}}$ \citep{Marty2012EPSL}.  The primitive mantle could have potentially contained even more, as much as 10--50 Earth oceans \citep{Dreibus1989OEPSA, Righter1999EPSL, Abe2000OEM}, although there is no definitive constraint on this value.  The core may also contain significant water, although that too is unclear.  \citet{Nomura2014Sci} estimate a quantity of hydrogen in the core equivalent to 80 Earth oceans of water, whereas \citet{Badro2014PNAS} find very little.  From this, a reasonable lower limit for the amount of water that must be delivered to (and retained by) the Earth is $5\times10^{-4}$ $\mathrm{M_{\earth}}$, although the actual amount of water delivered could be significantly larger.

\begin{figure}[t!]
\centering
\includegraphics[width=4.75in]{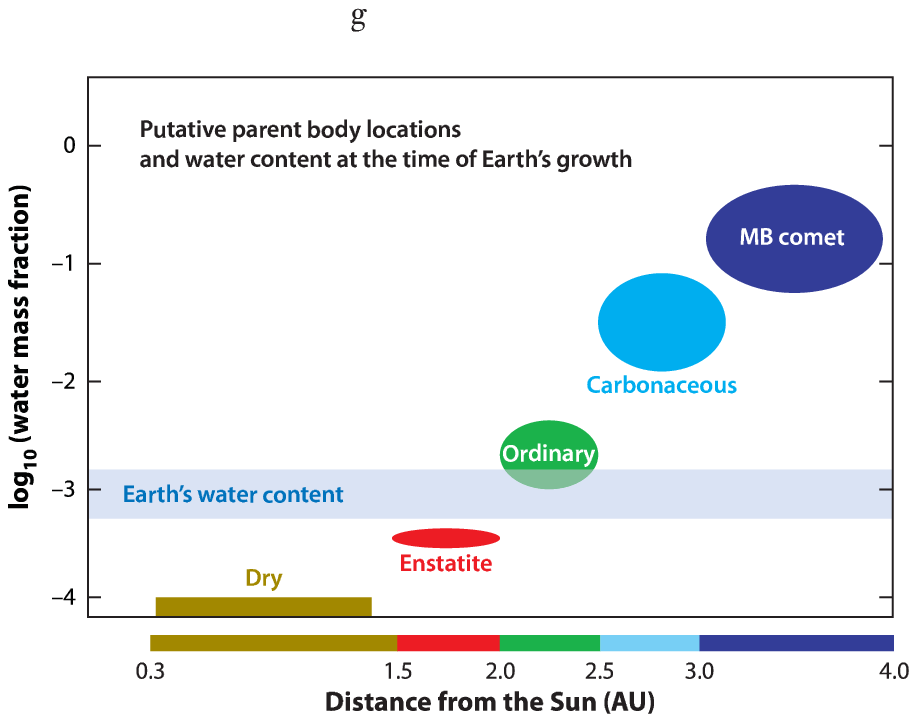}
\vspace*{0.1in}
\caption{Water contents of material estimated to have formed at different heliocentric distances, with the range of Earth's possible water content shown for comparison.  Ordinary chondrites are spectroscopically linked to S-type asteroids, which dominate the inner asteroid belt, while carbonaceous chondrites are thought to originate from C-type asteroids, which dominate the outer belt.  There is not a clear spectral link between enstatite chondrites and a known class of asteroids, although their chemistry suggests that they formed interior to the asteroid belt.  Figure from \citet{Morbidelli2012AREPS}.  \label{fig:watercont}}
\end{figure}

It has traditionally been assumed that the solar nebula around 1 AU was too hot for ice to condense and be incorporated into local Earth-forming material.  This is based in part on associations between different meteorite types and their asteroidal parent bodies or estimated locations of formation (Figure \ref{fig:watercont}), showing a trend towards extremely dry material near Earth's location.  However, nebula models taking into account stellar irradiation, viscous heating, and time-dependent opacities \citep[e.g.][]{Hueso2005AA, Davis2005ApJ, Garaud2007ApJ, Oka2011ApJ, Bitsch2015AA, Baillie2015AA} show that the ice-condensation line (the \textit{snowline}) can move inwards to $\sim$1 AU as the accretion of gas onto the central star slows.  \citet{Morbidelli2016Icar} show that even if the temperature around 1 AU is cold enough for ice to condense, the snowline could be fossilized at around 3 AU since the gas drifts inwards more quickly than the snowline moves, and the formation of a proto-Jupiter would block the migration of icy particles into the inner disk.  Thus, there is still a problem of condensing ice and incorporating it into planetesimals in the terrestrial planet zone.

Water is clearly present in bodies in the outer asteroid belt and beyond, however.  Primitive meteorites such as CM and CI chondrites can have $\sim$5-20\% water by mass, and are believed to come from asteroids such as the C-types that dominate the outer asteroid belt \citep{Burbine2002Ast3}.  Ceres at $\sim$2.7 AU contains several tens of percent water ice \citep[e.g.][]{McCord2005JGR} and shows signs of water vapor emission \citep{Kuppers2014Nat} and recent aqueous activity \citep{deSanctis2016Nat}.  Water ice has been directly detected on the asteroid Themis at $\sim$3.1 AU \citep{Campins2010Nat, Rivkin2010Nat}.  Some outer asteroid belt bodies called main-belt comets, which show signs of activity, likely have at least 10\% water ice \citep{Jewitt2012AJ}.  Centaurs, comets, trans-Neptunian objects, and many outer planet satellites are also ice-rich.

A key constraint on the origin of Earth's water is the deuterium/hydrogen (D/H) ratio, estimated by \citet{Lecuyer1998ChemGeo} to be $\sim150\times10^{-6}$, or about 6x the solar value.  This value is close to the mean whole-rock value for water-rich carbonaceous chondrite meteorites \citep[e.g.][]{Dauphas2000Icar, Robert2003SSRv, Robert2006MESS2, Alexander2012Sci}, suggesting that material similar to carbonaceous chondrites is a potential source of Earth's water.  The value for comets is often quoted as being 12x the solar value, or 2x terrestrial, although there are some exceptions to this, as described further in Section \ref{sec:comets}.  While carbonaceous chondrites may provide a good match to the Earth's D/H ratio, their oxygen isotope ratios and volatile element compositions, when compared to the Earth, suggest that carbonaceous chondritic material may have only contributed $\sim$2\% of the Earth's mass \citep{Drake2002Nat, Marty2012EPSL}.

There are geochemical constraints on the timing of water delivery as well.  Studies of core formation in the Earth suggest that the early material that accreted to form Earth must have been highly reduced \citep{Wood2008GCA, Rubie2011EPSL}, which would preclude the presence or delivery of significant water early on.  As we discuss in Section \ref{sec:latedeliv}, \citet{Rubie2015Icar} combined a core-mantle differentiation model with N-body simulations to explore this issue in further detail.

Some models \citep[e.g.][]{Rubie2004Nat, Siebert2013Sci} have proposed that core-mantle differentiation occurs under oxidizing conditions, which could be consistent with abundant water being accreted to Earth very early. In those models, accreted silicate material has a high FeO content (e.g. FeO mole fraction $X_{\mathrm{FeO}} \approx 0.2$). During core formation, oxygen partitions into the core (as FeO) and reduces $X_{\mathrm{FeO}}$ of the mantle to the observed value of $\sim$0.06. However, there is a significant problem with such models because several weight percent Si also partitions into the core, and every mole of Si that enters the core causes two moles of FeO to be transferred to the mantle.  This makes it impossible for the silicate mantle to achieve the observed low final $X_{\mathrm{FeO}}$ value \citep{Rubie2015Icar, Rubie2015TOG}.

\section{Stages of Terrestrial Planet Formation}
\label{sec:stages}

The process of terrestrial planet formation begins with the condensation of the first solids $\sim$4.567--4.568 Gyr ago \citep[e.g.][]{Amelin2002Sci, Bouvier2010NatGSci, Connelly2012Sci} and can be considered complete by the end of the heavy bombardment of the inner Solar System around 4.1--3.8 Gyr ago \citep{Tera1974EPSL, Chapman2007Icar, Bottke2012Nat, Marchi2012EPSL, Morbidelli2012EPSL}.  The gaseous solar nebula dissipated within $\sim$2--10 Myr of the formation of the first solids \citep[e.g.][]{Haisch2001ApJ, Kita2005ASPC},  and during that time, the initial planetesimals and planetary embryos formed and the giant planets grew their cores and captured their gaseous envelopes.  While Mars may essentially be a leftover embryo from this early phase \citep{Dauphas2011Nat}, the larger terrestrial planets experienced a much more drawn-out accretion process, lasting around 30--100 Myr \citep{Raymond2006Icar, Raymond2009Icar, Kenyon2006bAJ, OBrien2006bIcar, Touboul2007Nat, Kleine2009GCA, Morbidelli2012AREPS}.  For the Earth, the last major impact was likely that which formed the Moon \citep{Cameron2000OEM, Canup2001Nat, Canup2004Icar, Cuk2012Sci, Canup2012Sci, Reufer2012Icar}.  Here we give a brief overview of the different stages of terrestrial planet formation to provide a background for the rest of the chapter.  There have been numerous reviews of terrestrial planet formation over the last decade or so, which the reader can consult for further details \citep[e.g.][]{Chambers2004EPSL, Chambers2010Dust, Righter2011PNAS, Morbidelli2012AREPS, Raymond2014PP6}.

The formation of terrestrial planets has often been described as consisting of three main stages, although there may be some overlap between them.  The first stage, in which solid grains in the nebula accumulate into kilometer-scale or larger planetesimals is one of the most uncertain.  The main problem in this stage is that the radial drift of bodies in the nebula due gas drag effects \citep[e.g.][]{Weidenschilling1977MNRAS, Weidenschilling1980Icar} is fastest for meter-scale bodies, which are too large to be coupled to the gas, but small enough in mass that they feel a substantial drag force.  Bodies of this size can drift into the Sun on a timescale $\sim$100 years, and thus any planetesimal growth process must be rapid enough to overcome this \textit{meter size barrier}.

One proposal is that planetesimals grow by pairwise accretion of smaller grains in the nebula \citep[e.g.][]{Weidenschilling1980Icar, Weidenschilling1993PP3,  Weidenschilling1997Icar, Wurm2001Icar, Weidenschilling2011Icar}.  One potential issue with this scenario is that it requires a non-turbulent nebula, at least at the location and time when planetesimals were forming, which may not necessarily be the case \citep{Armitage2011ARAA}.  Otherwise, growth is slowed and bodies may not grow past the meter size barrier fast enough, and turbulence may drive collisions and erosion of bodies, rather than growth.

Another scenario is that particles settling to the midplane of the nebula may exceed a critical density such that planetesimals could grow purely by self-gravity, a process called \textit{gravitational instability} \citep[e.g.][]{Goldreich1973ApJ, Ward2000OEM, Youdin2002ApJ}.  Like with pairwise accretion, nebular turbulence can impede this process, but even if the nebula is initially non-turbulent, shear between the midplane particle layer and the overlying gas could drive turbulence and prevent the solids from reaching a critical density \citep[e.g.][]{Weidenschilling1993PP3, Weidenschilling2003Icar}.

Some models, however, take advantage of turbulence to aid in the formation of planetesimals.  In a process called \textit{turbulent concentration}, eddies in the solar nebula may preferentially concentrate bodies of specific sizes, in particular chondrule-sized bodies, and lead to rapid growth \citep{Cuzzi2001ApJ, Cuzzi2008ApJ, Chambers2010Icar}.  Alternatively, meter-scale boulders may become concentrated in over-dense regions of the disk, and the collective drag force on the solids forces the gas to move along with them (rather than sub-Keplerian).  This reduces the headwind and radial drift in a process called the \textit{streaming instability} \citep{Youdin2005ApJ, Johansen2007Nat}, and solids drifting into this region continue to accumulate until a critical density is reached.

Regardless of the specific process by which planetesimals form, they soon enter a second stage of growth in which they begin to gravitationally accrete one-another and grow into larger planetary embryos.  At first, this proceeds through a process called \textit{runaway accretion} \citep[e.g.][]{Greenberg1978Icar, Wetherill1989Icar, Kokubo1996Icar, Weidenschilling1997Icar}, in which bodies that first manage to grow larger are able to accrete even more rapidly due to their enhanced cross sections and gravitational focusing.  Once the growing embryos are large enough to significantly excite the relative velocities of surrounding planetesimals, their accretion rate slows and they enter a phase of \textit{oligarchic growth} in which other, smaller embryos are able to catch up \citep{Kokubo1998Icar, Kokubo2000Icar}.  The end result of this stage is a set of Lunar- to Mars-mass embryos embedded in a swarm of remnant planetesimals.

Recently, \textit{pebble accretion} has emerged as an extremely efficient process for the growth of planetary embryos and giant planet cores \citep[e.g.][]{Ormel2010AA, Lambrechts2012AA, Lambrechts2014AA, Morbidelli2012AA}.  So-called pebbles, defined as having a stopping time due to gas drag on the order of their orbital period (generally mm- to cm-scale), are efficiently accreted onto large planetesimals due to gas dynamic effects, and those planetesimals can rapidly grow into planetary embryos.  \citet{Levison2015PNAS} find that the terrestrial planets themselves can grow largely by accreting pebbles.  Water delivery has yet to be quantified in such a model.  Other models find that pebble accretion, at least in the terrestrial planet region, results in a swarm of embryos and remnant planetesimals much like the end result of runaway and oligarchic growth \citep{Jacobson2015GMS, Morbidelli2015Icar, Chambers2016ApJ}.

\begin{figure}[t!]
\centering
\includegraphics[width=5.0in]{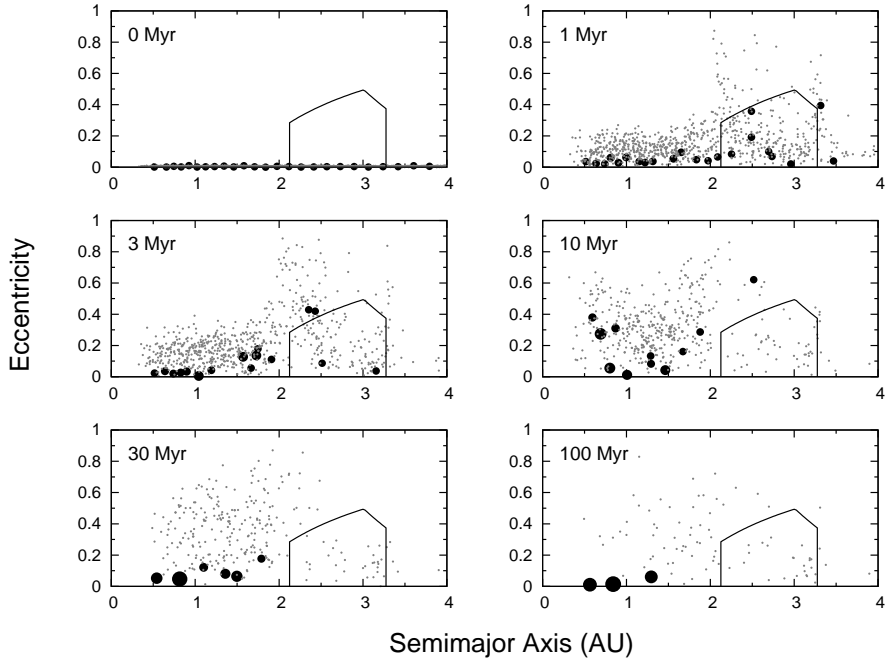}
\vspace*{-0.1in}
\caption{Terrestrial planet formation from the EJS1 simulation of \citet{OBrien2006bIcar}.  Large black dots are Mars-mass embryos, and small grey dots are planetesimal with 1/40 the mass of the embryos.  The outlined region from $\sim$2-3 AU is the asteroid belt.  Figure from \citet{OBrien2011SSRv}.  \label{fig:tpform1}}
\end{figure}

In the final stage of terrestrial planet formation, embryos collide violently with one-another and grow into the final planets on a timescale of $\sim$30-100 Myr.  In the classical scenario, embryos and planetesimals are spread through the terrestrial planet region, and likely into the asteroid belt zone, and the giant planets are fully-formed and located at or near their present locations \citep[e.g.][]{Chambers2001Icar, Raymond2004Icar, OBrien2006bIcar, Raymond2009Icar, Fischer2014EPSL}.  Figure \ref{fig:tpform1} shows the results of a simulation from \citet{OBrien2006bIcar}, starting with 25 Mars-mass embryos (large black dots) and 1000 planetesimals, each with 1/40 the mass of the embryos (small grey dots).  The gas disk is assumed to have dissipated by this point, so gas effects are not included.  While there is no gas to provide a damping effect on the growing planets, the swarm of planetesimals provides a damping effect due to \textit{dynamical friction} \citep[e.g.][]{Wetherill1993Icar}.  Terrestrial planets grow on a timescale of $\sim$100 Myr, and the asteroid belt region, outlined in the figure, is dynamically excited and depleted by perturbations from the giant planets and the mutual scattering of embryos \citep{Wetherill1992Icar, Petit2001Icar, OBrien2007Icar}.  An important aspect of this scenario from the point of view of water delivery is that material from beyond $\sim$2.5 AU can be scattered into the terrestrial planet region and accreted by the growing planets.  This will be discussed in further detail in Section \ref{sec:latedeliv}.

A major problem with this scenario is that the Mars analogues forming around 1.5 AU are typically 5-10 times more massive than Mars itself.  This problem of explaining the small mass of Mars was first noted by \citet{Wetherill1991LPSC} and persists over a wide range of parameter space, as shown by \citet{Raymond2009Icar}, \citet{Morishima2010Icar}, and \citet{Izidoro2015MNRAS}.  Numerous models have been tested in an attempt to reconcile this issue, as well as to fit the full range of constraints on terrestrial planet formation and Solar System evolution \citep[e.g.][]{Nagasawa2005ApJ, Thommes2008ApJ, Kominami2002Icar, Kominami2004Icar, Ogihara2007Icar, Morishima2008ApJ, Morishima2010Icar, Lykawka2013ApJ, Izidoro2014ApJ}.  Often these models have included the effects of a gas disk, which can provide eccentricity and inclination damping as well drive the migration of embryos through a process called \textit{Type 1 migration} \citep[e.g.][]{Goldreich1980ApJ, Ward1986Icar}.  In general, though, such models are still not able to fit the full range of constraints, or require initial conditions such as eccentric giant planet orbits that are not consistent with Solar System formation models.  For purposes of discussing water delivery in a classical scenario in Section \ref{sec:latedeliv}, we will use the \citet{OBrien2006bIcar} simulations described in the previous paragraph.  For a more thorough review of the other scenarios listed above, see \citet{Morbidelli2012AREPS} and \citet{Raymond2014PP6}.

A significant breakthrough in resolving the problem of forming a small Mars came when \citet{Hansen2009ApJ} showed that the mass distribution of the terrestrial planets can be reproduced if the initial distribution of material is confined to a narrow annulus between $\sim$0.7-1 AU.  While they did not propose a specific mechanism for achieving this initial distribution, their work led \citet{Walsh2011Nat} to model how the migration of the giant planets in the presence of nebular gas may allow for the truncation of the planetesimal/embryo distribution around 1 AU\footnote{Several mechanisms have subsequently been proposed to explain the inner edge at $\sim$0.7 AU, including a fossilized silicate condensation line \citep{Morbidelli2016Icar} and the outward migration of Jupiter's core in a model where it initially forms close to the Sun \citep{Raymond2016MNRAS}.}.  A large planet like Jupiter can open a gap in the nebular disk and migrate inwards via a process called \textit{Type 2 migration}, where the planet follows the viscous evolution of the disk \citep{Lin1986ApJ, Kley2012ARAA}.  If a second, smaller planet (e.g.~Saturn) is present just exterior to the first planet, however, the profile of the gap and the resulting torque balance changes, and the inward migration can be halted and reversed \citep[e.g.][]{Masset2001MNRAS, Morbidelli2007bIcar, Pierens2008AA, Pierens2011AA, Pierens2014ApJ, Zhang2010bApJ, DAngelo2012ApJ}.  This allows for a scenario in which Jupiter migrates into the terrestrial planet region, Saturn catches up with it, and the two planets open a mutual gap in the disk, reversing the direction of migration and moving Jupiter and Saturn back outwards to near their current locations.  \citet{Walsh2011Nat} termed this migration reversal the `Grand Tack', in analogy with the turning of a sailboat. 

\begin{figure}[t!]
\centering
\includegraphics[width=4.75in]{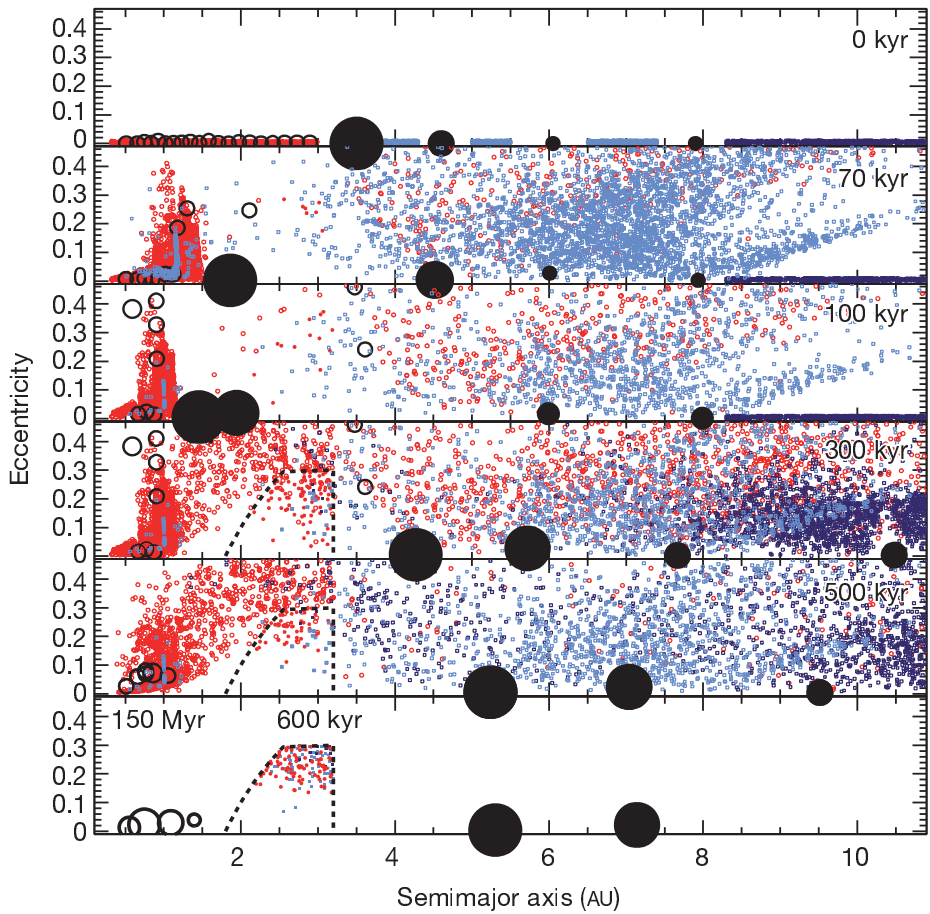}
\vspace*{0.2in}
\caption{Figure from from \citet{Walsh2011Nat} showing a simulation of the Grand Tack scenario.  The top three panels show the inward migration phase of the giant planets (lasting 100 kyr), and the last three panels show their outward migration (lasting 500 kyr).  The last panel also shows the terrestrial planets formed after 150 Myr.  Material implanted into the asteroid belt is not uniformly distributed, the inner asteroid belt is dominated by material originating within a few AU of the Sun (red dots), and the outer belt is dominated by material scattered inwards from the giant planet region (blue dots), which broadly reproduces the S-Type/C-Type dichotomy seen in the asteroid belt today \citep[e.g.][]{Gradie1982Sci, MotheDiniz2003Icar}.  The eccentricity distribution of asteroids after 600 kyr is higher than the observed distribution, but \citet{Deienno2016Icar} show that it evolves to roughly match the current distribution over the subsequent 4.5 Gyr of evolution.  The terrestrial planets will accrete some of the material scattered inwards from the giant planet region, which is likely water rich.  This was quantified in detail by \citet{OBrien2014Icar}. \label{fig:tpform2}}
\end{figure}

\citet{Walsh2011Nat} found that if Jupiter reversed direction at 1.5 AU, the inner disk would be truncated at $\sim$1 AU, and Mars would generally form from one or a few embryos that scattered outwards and became relatively isolated from the rest of the disk, as in the simulations of \citet{Hansen2009ApJ}.  Furthermore, while the passage of Jupiter through the primordial asteroid belt would be highly disruptive to the initial distribution of material there, \citet{Walsh2011Nat} showed that the scattering effects of Jupiter as it moved inwards and outwards actually repopulated the asteroid belt with original material from the asteroid belt region as well as material scattered inwards from the giant planet region.  Figure \ref{fig:tpform2} shows a simulation from \citet{Walsh2011Nat} that illustrates this process, with the top three panels showing the inward migration phase (over 100 kyr), and the last three showing outward migration (over 500 kyr).  The resulting inner asteroid belt is dominated by material originating within a few AU of the Sun (red dots), while the outer belt is dominated by material scattered inwards from the giant planet region (blue dots), broadly reproducing the S-Type/C-Type dichotomy seen in the asteroid belt today \citep[e.g.][]{Gradie1982Sci, MotheDiniz2003Icar}.  The eccentricity distribution of asteroids after 600 kyr is higher than the observed distribution, but will evolve to roughly match the current distribution over the subsequent 4.5 Gyr of evolution \citep{Deienno2016Icar}.  Material scattered inwards from the giant planet region would likely be water-rich and some would be accreted by the terrestrial planets, which was quantified by \citet{OBrien2014Icar} and will be discussed in further detail in Section \ref{sec:latedeliv}.  \citet{Brasser2016ApJ} performed more detailed simulations of the Grand Tack scenario and showed that a reversal of Jupiter at $\sim$2 AU might provide an even better match to some constraints than the 1.5 AU value used by \citet{Walsh2011Nat}, although the results in terms of water delivery are expected to be fairly similar.

More recently, there have been several models published besides the Grand Tack that can match many aspects of the inner Solar System. The `low-mass asteroid belt model' proposes that only a small population of planetesimals formed beyond 1-1.5 AU \citep{Izidoro2014ApJ, Izidoro2015MNRAS, Drazkowska2016AA, Morbidelli2016JGR, Raymond2017SciAdv}.  This is consistent with some models of planetesimal formation by the streaming instability in evolving disks \citep{Drazkowska2016AA, Surville2016ApJ}. This mass deficit naturally reproduces the large Earth/Mars mass ratio much like \citet{Hansen2009ApJ} and the Grand Tack, but requires a mechanism like chaotic excitation to match the asteroid belt \citep{Izidoro2016ApJ}.  In addition, some models based on pebble accretion can match the terrestrial planets and give a depleted asteroid belt \citep{Levison2015PNAS}.

\section{In-Situ/Early Water Delivery}
\label{sec:insitu}

Numerous studies have shown that water could be incorporated into olivine grains around 1 AU through adsorption directly from the gaseous nebula \citep{Stimpfl2006JCG, Muralidharan2008Icar,King2010EPSL, Asaduzzaman2015MAPS}.  These grains would then follow the steps outlined in Section \ref{sec:stages} and grow into planetesimals, potentially contributing several oceans worth of water to a growing Earth from local, early-accreted material.  Several key questions remain, however.  Studies show that the early material that accreted to form Earth was highly reduced \citep[e.g.][]{Wood2008GCA, Rubie2011EPSL}, which would not be the case if it contained significant water.  Also, enstatite chondrite meteorites are drier than the Earth (see Figure \ref{fig:watercont}) and highly reduced, and may be difficult to explain if significant water-bearing material was present.  An additional criticism is that this process would lead to the capture of water with a D/H ratio close to Solar, whereas the D/H ratio of the Earth's water is approximately 6x Solar. \citet{Ganguly2016MetSoc} show that a fractionation process between adsorbed and nebular water may explain at least part of this enrichment.

Another possibility for the early incorporation of water-bearing material is the oxidation of an early hydrogen atmosphere by FeO in the terrestrial magma ocean \citep{Ikoma2006ApJ, Genda2008Icar}.  This requires fast growth of the planet in order to capture hydrogen from the solar nebula, which likely dissipated within 10 Myr \citep{Haisch2001ApJ, Kita2005ASPC}, in contrast to radiometric estimates for the formation timescale of the Earth that are much longer \cite[e.g.][]{Touboul2007Nat, Kleine2009GCA}.  The D/H ratio of the initial captured gas would be solar, not terrestrial.  Hydrodynamic escape of atmosphere over billions of years could potentially lead to a terrestrial D/H ratio \citep{Genda2008Icar}.  However, this timescale is much longer than the $\sim$100 Myr closure age of the atmosphere estimated from I-Xe isotopes \citep{Wetherill1975ARNPS, Avice2014RSPT}.

Even if these two processes are not able to explain all of the Earth's water, or give a terrestrial D/H ratio, they may have still occurred to some degree.  Deep-mantle lavas sampling relatively unmixed reservoirs show evidence for solar D/H \citep{Hallis2015Sci}, suggesting that at least some of Earth's water could have been incorporated directly from the solar nebula.

Finally, \citet{Ciesla2005EPSL} proposed that water-bearing phyllosilicate dust could drift inwards and be incorporated into planetesimals around 1 AU.
This would require that water be locked in phyllosilicates while nebular gas was still around.  However, many phyllosilicates found in meteorites seem to have been formed by parent-body processes after nebular gas would have dissipated.

\section{Cometary Delivery}
\label{sec:comets}

Comets have long been suggested as a source of terrestrial water, given their obvious ice-rich nature \citep[e.g.][]{Chyba1987Nat, Delsemme1992ASR, Delsemme1997Comets, Delsemme1998PSS, Owen1995Icar}.  The most common objection to this is that most comets with measured D/H ratios have about twice the terrestrial value.  Initially all of those measured comets were Oort-cloud or Halley-type comets.  The Jupiter-family comet 103P/Hartley 2 was recently found to have a terrestrial D/H ratio \citep{Hartogh2011Nat}, as was 45P/Honda-Mrkos-Pajdu{\v s}{\'a}kov{\'a} \citep{Lis2013ApJ}, hinting that JFCs could be a possible source.  However, comet 67P/Churyumov-Gerasimenko, also a Jupiter-family comet, was found to have a D/H nearly 3 times terrestrial \citep{Altwegg2015Sci}, so there is no clear trend from the current observational data.

Possibly more problematic than any geochemical constraints, however, is the significant dynamical difficulty in delivering the Earth's water primarily through comets.  The probability that comets originating in the giant planet region hit the Earth is $\sim 1\times10^{-6}$ \citep{Morbidelli2000MAPS}.  Various models of outer Solar System evolution suggest that $\sim$30-50 $\mathrm{M_{\earth}}$ of material may have been present in a primordial cometessimal disk \citep[e.g.][]{Malhotra1993Nat, Malhotra1995AJ, Hahn1999AJ, Gomes2004Icar, Gomes2005Nat, Tsiganis2005Nat, Morbidelli2010AJ}.  Thus, if 50 $\mathrm{M_{\earth}}$ of comets were initially present and scattered by the giant planets, $5\times10^{-5}$ $\mathrm{M_{\earth}}$ of cometary material would have hit the Earth.  As discussed in Section \ref{sec:constraints}, the Earth's water content is at least $5\times10^{-4}$ $\mathrm{M_{\earth}}$, so even if those comets were pure ice, only $\sim$10\% of Earth's water could be provided by a cometary source.

\section{Later Delivery}
\label{sec:latedeliv}

\citet{Morbidelli2000MAPS} and \citet{Raymond2004Icar} showed that if the initial disk of planetesimals and embryos in the inner Solar System extended through the asteroid belt region, water-bearing material from beyond $\sim$2.5 AU can be incorporated into the growing terrestrial planets, often relatively late in the accretion process.  \citet{OBrien2006bIcar} performed several sets of simulations with higher resolution that allowed for the analysis of this delivery mechanism in greater detail \citep[see also][]{Raymond2006Icar, Raymond2007AsBio}.  The \citet{OBrien2006bIcar} simulations started with 25 Mars-mass embryos and 1000 planetesimals, each with 1/40 the mass of an embryo (so that there is equal mass in the planetesimal and embryo populations).  Two configurations of Jupiter and Saturn were used: The EJS simulations, with Jupiter and Saturn on their current, eccentric orbits, and the CJS simulations, with Jupiter and Saturn on more circular and closely-spaced orbits, as predicted by the Nice Model \citep{Tsiganis2005Nat,Gomes2005Nat,Morbidelli2005Nat}.  Four simulations were run for each giant planet configuration.

\begin{figure}[p!]
\centering
\vspace*{0.4in}
\includegraphics[width=4.75in]{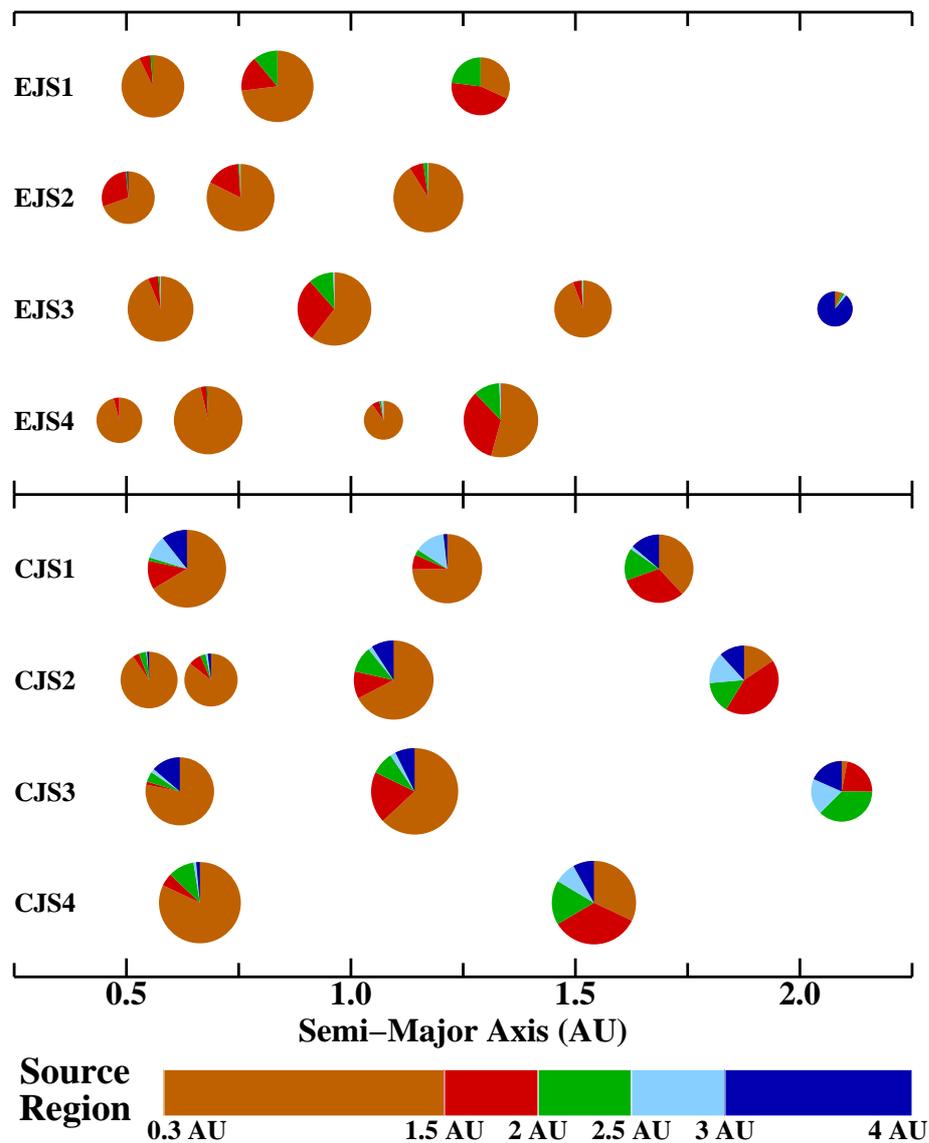}
\vspace*{0.3in}
\caption{Relative contributions of material from different semimajor axis zones in the simulations of \citet{OBrien2006bIcar} (adapted from a figure in that paper). The diameter of each symbol is proportional to the diameter of the planet, assuming that they all have the same density.  In the EJS simulations, Jupiter and Saturn are on their current, eccentric orbits, and in the CJS simulations, Jupiter and Saturn are on more circular and closely-spaced orbits, as predicted by the Nice Model \citep{Tsiganis2005Nat,Gomes2005Nat,Morbidelli2005Nat}. \label{fig:cjsejs}}
\end{figure}

Figure~\ref{fig:cjsejs} shows the final planets in the simulations of \citet{OBrien2006bIcar}, with the contributions of material from different semimajor axis regions.  12 terrestrial planets were formed in the four CJS simulations.  All of them accreted at least a few planetesimals from beyond 2.5 AU (median number of 7), and 9 of the planets contained at least one embryo that originated from beyond 2.5 AU.  The fraction of each planet's total mass that originates from beyond 2.5 AU ranges from 1.6\% to 38\%, with a median value of 15\%.  On average, about 85\% of the material from beyond 2.5 AU that is accreted by the planets is delivered by embryos.

In contrast, the 14 planets in the EJS simulations receive very little material from beyond 2.5 AU.  None of the planets accrete an embryo from beyond 2.5 AU, although one planet (in simulation EJS3) originated as an embryo beyond 2.5 AU and never accreted any other embryos.  At most, planets accrete 3 planetesimals from beyond 2.5 AU, and 5 of the planets contain no material at all from beyond 2.5 AU.  The fraction of each planet's total mass that originates from beyond 2.5 AU (excluding the single-embryo planet in EJS3) has a median value of 0.3\% and ranges from 0\% to 1.6\%.  This profound disparity between the two sets of simulations is primarily due to the fact that resonances in the asteroid belt region are stronger with Jupiter and Saturn on eccentric orbits, and this leads to most material from beyond 2.5 AU being driven into the Sun or ejected from the Solar System, whereas more of that material slowly diffuses into the terrestrial planet region in the CJS simulations.  Interestingly, earlier simulations incorporating only embryos \citep{Morbidelli2000MAPS, Raymond2004Icar} found that sufficient water could be delivered to the terrestrial planets if Jupiter and Saturn are eccentric.  In the \citet{OBrien2006bIcar} simulations, the large number of planetesimals tends to aid in the clearing of embryos from the region beyond 2.5 AU, such that little or no material from that region ends up being incorporated into the planets when Jupiter and Saturn are eccentric.

The mass fractions of material from beyond 2.5 AU in the final planets can be converted to estimates of water content, for comparison to the Earth.  As noted above, the median mass fraction for planets in the CJS simulations is 15\% (ranging from 1.6-38\%).  Assuming a mass fraction of 10\% water for the material originating from outside 2.5 AU, consistent with the values for carbonaceous chondrites, and also assuming that all accreted volatiles are retained by the planet and not lost in the impact, a 15\% mass fraction implies 0.015 $\mathrm{M_{\earth}}$ of water for an Earth-mass planet, or 30$\times$ the lower limit of $5\times10^{-4}$ $\mathrm{M_{\earth}}$ from Section \ref{sec:constraints}.  The full range of values for the planets spans 3-75$\times$ the lower limit.  Even assuming a more conservative mass fraction of water in the material beyond 2.5 AU of 5\% and assuming that only 10\% of the water is retained in the impact, the median amount of water in an Earth-mass planet is still 1.5$\times$ the lower limit.  We note that if the material from beyond 2.5 AU is similar to carbonaceous chondrites, the large amount delivered in these simulations could pose a geochemical conflict.  As discussed in Section \ref{sec:constraints}, \citet{Drake2002Nat} and \citet{Marty2012EPSL} argue that, based on oxygen isotope ratios and volatile element abundances, carbonaceous chondrites may have only contributed about 2\% of the Earth's mass.

For the EJS simulations, the median mass fraction of material from beyond 2.5 AU that ends up in the final terrestrial planets is 0.3\%, with a maximum value of 1.6\%.  Assuming a 10\% mass fraction of water in the impactors from beyond 2.5 AU, and 100\% retention of volatiles in impacts, the median water content for an Earth-mass planet would be $3\times10^{-4}$ $\mathrm{M_{\earth}}$, less than the lower limit value of  $5\times10^{-4}$ $\mathrm{M_{\earth}}$.  Hence, simulations with an eccentric Jupiter and Saturn have a much harder time forming Earth-like planets, at least in terms of water content.

\begin{figure}[t!]
\centering
\includegraphics[width=5.0in]{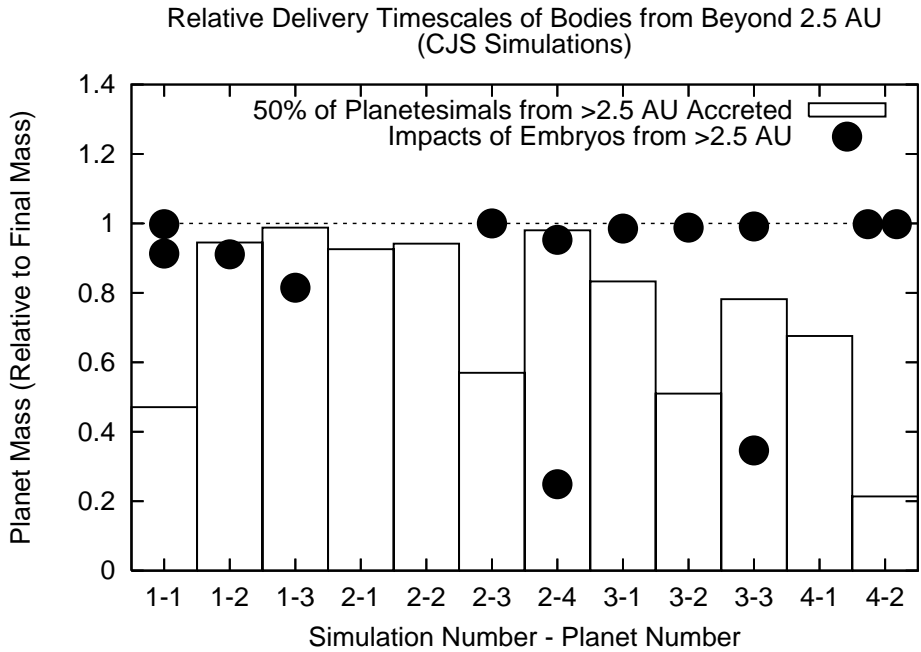}
\vspace*{0.1in}
\caption{Timing of the accretion of material from beyond 2.5 AU, which likely contains water and other volatiles, in the CJS simulations of \citet{OBrien2006bIcar}.  Each column is for a given planet in a given simulation.  The bars show the mass of each planet at the point when it has accreted 50\% of its final number of planetesimals originating from beyond 2.5 AU.  Almost all planets reach the 50\% point after they are more than half grown, implying that water-bearing planetesimals are delivered preferentially late rather than early.  Even more striking is the timing of accretion of embryos from beyond 2.5 AU.  The large black dots give the planet mass following the impact of these embryos.  Planet 4-2 is impacted by a body containing 2 embryos from beyond 2.5 AU, and only 3 planets do not experience the impact of an embryo from beyond 2.5 AU.  In most cases, the impact of an embryo from beyond 2.5 AU is the final embryo impact on a planet.  Planetesimals from beyond 2.5 AU that are `pre-accreted' by another embryo before hitting the planet are not counted in this graph.  Figure from \citet{OBrien2006bIcar}. \label{fig:wetimpacts}}
\end{figure}

\citet{OBrien2006bIcar} also analyzed the timing of the delivery of material from beyond 2.5 AU.  Figure \ref{fig:wetimpacts} shows the timing for the planets in the CJS set of simulations.  For each planet, the large black dots show the planet mass following the accretion of any embryos from beyond 2.5 AU, and the vertical bars show the planet mass when it has accreted 50\% of its final number of planetesimals originating from beyond 2.5 AU.  Nearly all of the embryo impacts occur well after the planet is half grown.  In fact, 8 of the 9 planets that do experience at least one impact by an embryo from beyond 2.5 AU actually have that embryo as their final large impactor.  Similarly, for the majority of planets, 50\% of their final number of planetesimals from beyond 2.5 AU are not delivered until after the planet has grown to half its size.  So whether from embryos or planetesimals, the delivery of potential volatile-carrying material tends to occur late in a planet's growth, and volatile-carrying embryos often occur as the final large impactor.    Given the fact that the planets in the EJS simulations contain no embryos from beyond 2.5 AU and at most three such planetesimals, a similar plot can not be constructed for the EJS simulations.  

Having water delivered relatively late in the accretion process makes it more likely to be retained, and is also consistent with the the findings of \citet{Wood2008GCA} and \citet{Rubie2011EPSL} that early material accreting to form Earth must have been highly reduced and the oxidation state increased with time.  We stress, though, that while water is delivered relatively late in these simulations, it is not delivered as a \textit{late veneer}, which is generally considered to be the mass delivered after differentiation has ceased.  Differentiation and core formation would almost entirely remove highly siderophile elements from the mantle, and the fact that some do exist today suggests that $\lesssim$1\% of Earth's total mass was accreted after differentiation ceased \citep{Drake2002Nat}.  \citet{OBrien2006bIcar} assumed that differentiation occurs as long as embryo impacts are occurring, and calculated the late veneer mass as that arriving after the final giant impact on a given planet.  For the CJS simulations, they find a median value of 1.2\% for planets that are comparable in mass to the Earth, reasonably consistent with the \citet{Drake2002Nat} estimate.  The median mass fraction of material from beyond 2.5 AU  that arrives after the last giant impact, as a fraction of planet mass, is 0.7\%, and ranges from 0 - 2.9\%, whereas the total median mass fraction of material from beyond 2.5 AU is 15\%.  Thus, the vast majority of water-bearing material arrives prior to the late veneer in these simulations.

As can be seen in Figure \ref{fig:cjsejs}, the simulations of \citet{OBrien2006bIcar} do not do a good job of reproducing Mars---even in the EJS simulations, the best Mars analogues are at least several times too massive.  With larger numbers of EJS and CJS simulations (50 each), \citet{Fischer2014EPSL} found that Mars-like planets could be produced in a small fraction of cases, but it is a very low-probability event, especially in the CJS case.  This is an issue for nearly all simulations of the classical scenario, except in the case where Jupiter is made to be substantially more eccentric than its current value \citep{Raymond2009Icar}.  This issue was part of the motivation for the work of \citet{Hansen2009ApJ} and the development of the Grand Tack model of \citet{Walsh2011Nat} described in Section \ref{sec:stages}.  \citet{OBrien2014Icar} extended the simulations of \citet{Walsh2011Nat} to investigate water delivery in the Grand Tack scenario in more detail.  Here we describe their key results.

The \citet{Walsh2011Nat} simulations began with a population of embryos with 1/2 or 1/4 of a Mars mass and a swarm of planetesimals extending out to 3 AU, and populations of primitive planetesimals between the four giant planets (the \textit{belts}) and beyond Neptune (the \textit{disk}).  The inward and outward migration of the giant planets pushes much of the inner Solar System material to inside of 1 AU, roughly reproducing the initial conditions of \citet{Hansen2009ApJ}, scatters material out of and then back into the asteroid belt, and scatters some of the material from the belt and disk populations into the terrestrial planet region as well.  \citet{Walsh2011Nat} did not explicitly track the accretion of the belt and disk material, but estimated that sufficient water would be delivered to explain the Earth's water budget.  \citet{OBrien2014Icar} took the simulations of \citet{Walsh2011Nat} immediately following the inward-then-outward migration of the giant planets and integrated them to 150 Myr, explicitly including the material originating from between and beyond the giant planets.  16 planetary systems were generated.

\citet{OBrien2014Icar} find that for planets larger than 0.75 $\mathrm{M_{\earth}}$ the median fraction of material originating from the belts population $f_{belts}$ is 2.3\%, and for the disk population $f_{disk}$ is 0.7\%.  Because the masses of the belt and disk populations were normalized such that each could individually explain the mass of C-type asteroids in the main belt, the total mass fraction of material from the giant planet region that ends up in the terrestrial planets is not the sum of $f_{belts}$ and $f_{disk}$, but rather a value between the two (i.e.~between 0.7 and 2.3\%).  Note that if the primitive planetesimals are similar to carbonaceous chondrites, these values are consistent with the estimates of \citet{Drake2002Nat} and \citet{Marty2012EPSL} that carbonaceous chondrites only contributed about 2\% of the Earth's total mass.

For the same assumption as made for the \citet{OBrien2006bIcar} simulations, that the primitive bodies have a water mass fraction of 10\% and no water is lost in collisions, Earth-mass planets would have a median water content of $2.3 \times10^{-3}$ $\mathrm{M_{\earth}}$ if the water were delivered entirely by the belts population or $7 \times10^{-4}$ $\mathrm{M_{\earth}}$ if it were delivered entirely by the disk population (and most likely a value somewhere between the two).  Both values exceed the lower limit of $5\times10^{-4}$ $\mathrm{M_{\earth}}$ discussed in Section \ref{sec:constraints} (4.6 and 1.4 times larger, respectively).  We note that the Grand Tack scenario has the important feature of matching both the water budget of Earth and the mass of C-type asteroids in the main belt in a self-consistent manner---a constraint that any model of water delivery by carbonaceous chondrite material must be able to match.

The 10\% value assumed above for the water content of primitive planetesimals originating from the giant planet region may be conservative, given that main-belt comets may have a much larger water content than this \citep{Jewitt2012AJ}. The discovery of water vapor emanating from Ceres \citep{Kuppers2014Nat} and the detection of water ice on the surface of Themis \citep{Campins2010Nat, Rivkin2010Nat} also suggest that the water content of primitive material may be larger than estimated from meteorites (in which only water bound to the silicates has survived).  Thus, larger water contents are a possibility, and would not be inconsistent with the Earth's total water budget since several tens of oceans worth of water can potentially be stored in the mantle, as discussed in Section \ref{sec:constraints}.  A higher planetesimal water content could also allow for significant impact-related losses of water, while still delivering the minimum required amount.

\begin{figure}[t!]
\centering
\begin{minipage}[b]{0.5\linewidth}
\centering
\includegraphics[width=2.35in]{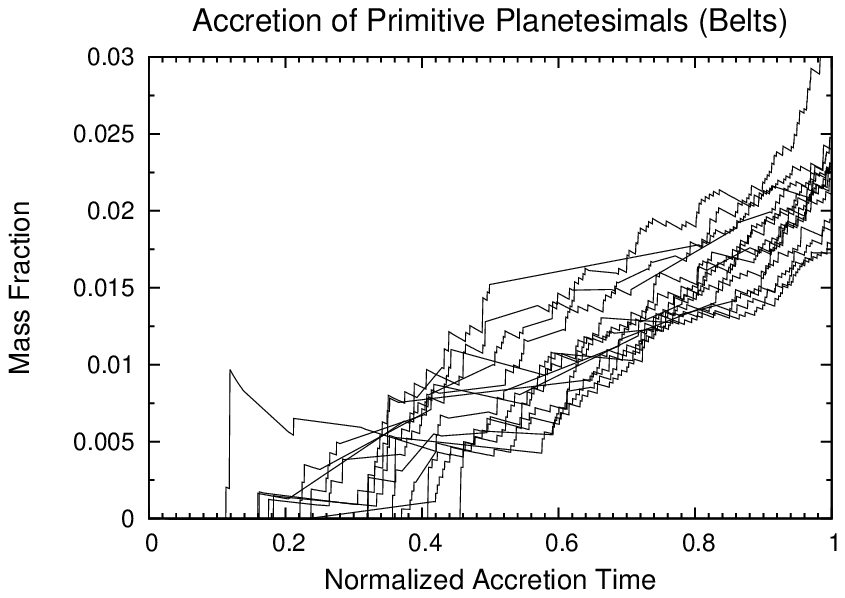}
\end{minipage}%
\begin{minipage}[b]{0.5\linewidth}
\centering
\includegraphics[width=2.35in]{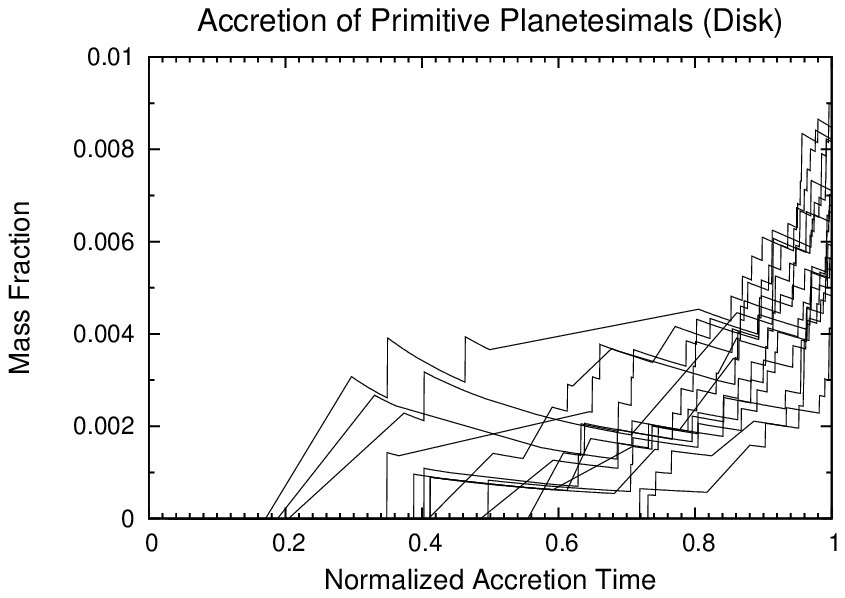}
\end{minipage}
\vspace*{0.1in}
\caption{Accretion curves for all planets larger than 0.75 $\mathrm{M_{\earth}}$ in the Grand Tack simulations of \citet{OBrien2014Icar}.  The mass fraction of primitive planetesimals from both the `belts' and `disk' populations are shown, where the belts population originally lies between the orbits of the giant planets, and the disk population initially lies beyond them.  Accreted fractions are plotted vs.~normalized time, which is the mass of the planet divided by its final mass.  The curves all increase with time, which shows that the primitive material, and hence water and other volatiles, arrives relatively late during the accretion process.  Figure from \citet{OBrien2014Icar}. \label{fig:gtaccret}}
\end{figure}

Figure \ref{fig:gtaccret} shows the timing of accretion of water-bearing material.  $f_{belts}$ and $f_{disk}$ are plotted vs.~time for all planets whose final mass is larger than 0.75 $\mathrm{M_{\earth}}$.  The time axis is in terms of \textit{normalized accretion time}, which is the mass of the planet as a fraction of its final mass.  If equal amounts of primitive material were accreted during each impact, all lines on the plot would be horizontal; the fact that the curves in Figure \ref{fig:gtaccret} all increase with increasing time means that the primitive planetesimals preferentially arrive late in the planets' accretion.  The delivery of primitive material late in the accretion process, when the planet is larger, means that water and volatiles that may be delivered by those impactors are more likely to be retained by the planet.  \citet{Wood2008GCA} and \citet{Rubie2011EPSL} show that material accreting to form Earth was initially highly reduced, and the oxidation state increased with time, consistent with late addition of water to the terrestrial planets.  \citet{Rubie2015Icar} integrated their multistage core-mantle differentiation model with actual Grand Tack simulations, and found that they can produce Earth-like planets with mantle concentrations of non-volatile elements and water that are close to the estimated values for the Earth's primitive mantle.

As with the CJS and EJS simulations described previously, the fact that primitive material is delivered late does not necessarily mean that water is all delivered as a late veneer at the tail end of accretion.  Water delivery is still a relatively gradual process, with the planets continuing to grow while accreting water-rich planetesimals.  The late veneer mass that is accreted in these simulations, however, tends to be significantly larger than in the CJS simulations, which were reasonably consistent with the estimate of $\lesssim$1\% total mass from \citet{Drake2002Nat}.  Planets larger than 0.75 $\mathrm{M_{\earth}}$ have a median late veneer mass (defined as mass accreted after the last embryo impact) of 18\%, only a fraction of which is primitive water-bearing material.  This high value is in part due the fact that many planets experience their last embryo impact in less than 20 Myr, and as shown by \citet{Jacobson2014Nat}, there tends to be an inverse correlation between late veneer mass and the time of last giant impact.  Of the 27 planets larger than 0.5 $\mathrm{M_{\earth}}$ formed in the \citet{OBrien2014Icar} simulations, only 8 have last giant impacts after 20 Myr and 5 of those have late veneer masses of a few percent or less.  \citet{Jacobson2014Nat} have shown that the timing of the last giant impact and the late veneer mass can be strongly affected by the initial embryo mass and embryo/planetesimal mass ratio.  This suggests a wider range of initial conditions should be explored to determine which parameters are able to best fit all available constraints.

\citet{Raymond2017Icar} recently proposed a new scenario for water delivery by planetesimals scattered inward during the giant planets' growth and migration, which has implications for the other models discussed above.  Jupiter and Saturn experienced a phase of runaway gas accretion during which the planets' masses increased dramatically, until they carved a gap in the disk and accretion slowed \citep[e.g.][]{Lissauer2009Icar}.  This growth was likely accompanied by orbital migration, although there are a number of viable migration histories \citep{Pierens2014ApJ}.  The giant planets' growth invariably destabilizes the orbits of nearby planetesimals, gravitationally scattering them in all directions onto eccentric orbits. Given the dissipative effects of aerodynamic gas drag \citep{Adachi1976PTP}, planetesimals' eccentricities may be damped, decoupling the planetesimals from the giant planets and preventing further scattering.  A fraction of planetesimals is naturally deposited interior to Jupiter's orbit, preferentially populating the outer part of the main asteroid belt \citep{Raymond2017Icar}.  Implanted planetesimals originate mainly between 4 and 10 AU (but with a tail out to 15-20 AU) and match the orbits of the C-type asteroids. In fact, Ceres is located immediately adjacent to the peak in the distribution of 1000 km-class planetesimals implanted by this mechanism \citep{Raymond2017Icar}.  Some planetesimals are scattered interior to the asteroid belt by Jupiter, crossing the orbits of the terrestrial planets.  This preferentially happens for large planetesimals and later in the disk lifetime, when gas drag is weaker.

The scenario proposed in \citet{Raymond2017Icar} would occur before any significant planetary growth in the terrestrial planet region, and hence could fit within any model of terrestrial planet formation.  In the Grand Tack model, for example, there would be an additional, earlier scattering of water-rich material into the terrestrial planet region by the \citet{Raymond2017Icar} mechanism, in addition to the later scattering of material during the inward and outward migration of Jupiter.  The amount of material scattered inwards during this early phase, however, is unconstrained since it is based on the (unknown) planetesimal distribution in the primordial Jupiter-Saturn zone.  Likewise, the timing of accretion of that material is unclear---even though it is implanted early, it may be accreted relatively late, especially if it is on eccentric orbits.  Future modeling of terrestrial planet formation incorporating the \citet{Raymond2017Icar} scenario, along with geochemical constraints \citep[e.g.][]{Rubie2015Icar}, may help better understand the magnitude and effects of this early planetesimal scattering process.

\section{Summary and Conclusions}
\label{sec:summary}

We have reviewed the main scenarios for delivery of water to the terrestrial planets.  Several early/in-situ mechanisms have been proposed, including the incorporation of water into olivine grains around 1 AU through adsorption directly from the gaseous nebula \citep{Stimpfl2006JCG, Muralidharan2008Icar,King2010EPSL, Asaduzzaman2015MAPS}, oxidation of an early hydrogen atmosphere by FeO in the terrestrial magma ocean \citep{Ikoma2006ApJ, Genda2008Icar}, and incorporation of inward-drifting phyllosilicate dust into planetesimals around 1 AU \citep{Ciesla2005EPSL}.  As discussed in Section \ref{sec:insitu}, each of these scenarios has its problems, for example potential inconsistencies with the measured D/H ratio of the Earth, and all run counter to work showing that the material that first accreted to form the Earth was highly reduced \citep{Wood2008GCA, Rubie2011EPSL}.  Cometary delivery has also been proposed, as discussed in Section \ref{sec:comets}.  It too may have issues matching the measured D/H of the Earth, but a bigger issue is that the delivery efficiency of comets is very low \citep{Morbidelli2000MAPS}, such that even delivering a single Earth ocean of water is unlikely.

Later-stage delivery of water-bearing material, in the course of planetary accretion, is a more promising scenario (Section \ref{sec:latedeliv}).  In the classical scenario, with planetesimals and embryos extending through the outer asteroid belt region, substantial amounts of water can be delivered to the terrestrial planets if Jupiter and Saturn are on circular, low-inclination orbits as predicted by the Nice Model \citep[e.g.][]{OBrien2006bIcar, Raymond2009Icar}.  However, it is possible that so much carbonaceous-chondrite-like material could be delivered in that case that the volatile element abundances and oxygen isotope ratios of the final planets are not consistent with Earth's values \citep{Drake2002Nat, Marty2012EPSL}.  Such simulations are also not able to reproduce the small mass of Mars.  In the Grand Tack scenario \citep{Walsh2011Nat, OBrien2014Icar}, where inward and outward migration of the giant planets truncates the terrestrial planetesimal/embryo distribution and scatters material inwards from the giant planet region, the final terrestrial planet systems are a better match to the actual Solar System (with a small Mars).  Sufficient water delivery occurs to explain the Earth's water budget without delivering excessive carbonaceous material that would violate the constraints of \citet{Drake2002Nat} and \citet{Marty2012EPSL} regarding volatile elements and oxygen isotope ratios.  Water is delivered relatively late in the accretion process, but not as a late veneer.  The later addition of water is consistent with evidence that the earliest material accreted to the Earth was highly reduced, and later material was more oxidized \citep{Wood2008GCA, Rubie2011EPSL, Rubie2015Icar}.  While some issues still remain with the Grand Tack scenario, specifically regarding the accretion timescales of the planets that form, this discrepancy may be reduced as a wider range of initial conditions is explored \citep[e.g.][]{Jacobson2014Nat}.  Finally, new models of terrestrial planet formation have been proposed since the introduction of the Grand Tack scenario, such as those based on pebble accretion \citep[eg.][]{Levison2015PNAS} or having an initially low mass in the asteroid belt region \citep{Izidoro2014ApJ, Izidoro2015MNRAS, Drazkowska2016AA, Morbidelli2016JGR, Raymond2017SciAdv}, as well as new mechanisms for water delivery, such as early inward scattering of planetesimals during the formation of the giant planets \citep{Raymond2017Icar}.  These may be promising avenues for future research.

\section*{Acknowledgements}

We thank the International Space Science Institute (ISSI) for organizing and supporting the workshop ``The Delivery of Water to Protoplanets, Planets and Satellites'', and an anonymous reviewer for their helpful comments and suggestions.  AI acknowledges financial support from FAPESP through grants number 16/12686-2 and 16/19556-7. SNR acknowledges Agence Nationale pour la Recherche grant ANR-13-BS05-0003-002 (grant MOJO).  DCR and SAJ were supported by the European Research Council Advanced Grant ``ACCRETE'' (contract number 290568), and additional support to DCR was provided by the German Science Foundation (DFG) Priority Programme SPP1833 ``Building a Habitable Earth'' (Ru1323/10-1).


\end{document}